\documentclass[twocolumn,article,superscriptaddress,groupedaddress]{revtex4}
\usepackage[export]{adjustbox}
\usepackage{booktabs}
\usepackage{units}
\usepackage{graphicx}
\usepackage{enumitem}
\usepackage{xcolor}
\usepackage{hyperref}
\usepackage{psfrag}
\usepackage{graphicx}
\usepackage{dcolumn}
\usepackage{bm}
\usepackage{textcomp}
\usepackage{amsmath}

\begin{document}


\title{Cryogenic thermo-acoustics in the SPIRAL2 LINAC}

\author{Adnan Ghribi}\email{ghribi@in2p3.fr}
\affiliation{Grand Acc\'el\'erateur National d'Ions Lourds (GANIL)}%
\affiliation{Centre National de la Recherche Scientifique (CNRS - IN2P3)}%
\author{Muhammad Aburas}%
\affiliation{Grand Acc\'el\'erateur National d'Ions Lourds (GANIL)}%
\affiliation{Commissariat à l'Energie Atomique (CEA - IRFU)}%
\author{Abdallah Alhaffar}%
\affiliation{Ecole Centrale Supélec}%
\affiliation{Grand Acc\'el\'erateur National d'Ions Lourds (GANIL)}%
\author{Pierre-Emmanuel Bernaudin}%
\affiliation{Grand Acc\'el\'erateur National d'Ions Lourds (GANIL)}%
\affiliation{Commissariat à l'Energie Atomique (CEA - IRFU)}%
\author{Patxi Duthil}%
\affiliation{Irène Joliot-Curie Laboratory (IJCLab)}%
\affiliation{Centre National de la Recherche Scientifique (CNRS - IN2P3)}%
\author{Maroun Nemer}%
\affiliation{Centre Efficacité Energétique des Systèmes (CES)}%
\affiliation{MINES ParisTech}%
\author{Jean-Pierre Thermeau}%
\affiliation{Laboratoire AstroParticule et Cosmologie (APC)}%
\affiliation{Centre National de la Recherche Scientifique (CNRS - IN2P3)}%
\date{\today}

\begin{abstract}
SPIRAL2 is a superconducting LINAC subject to cryogenic thermo-acoustic oscillations occurring in its valves-boxes. 4 years of monitoring and experimental investigations with thousands of datasets turned these unwanted effects into an opportunity to study and understand thermo-acoustics in a complex environment such as a real life accelerator. Without digging deep into Rott's thermo-acoustics theory, thoroughly shown in other works, this paper describes the instrumentation and the methods that prepare more advanced modelling of these phenomena either to damp or to harness the energy of cryogenic thermo-acoustics.

\end{abstract}

\maketitle

\section{\label{sec:1}Introduction}
Superconducting linear accelerators (LINACs) provide electron and light and heavy ion beams for a wide
range of applications ranging from nuclear and atomic physics to
health and solid-state physics. SPIRAL2 is one of them
\cite{Lewitowicz:2006fx, ferdinand:in2p3-00867502,
  dolegieviez:hal-02187926}. It provides some of the most intense ion
beams to study the structure of the atoms and the nuclei. The recent
commissioning of the LINAC highlighted some phenomena known to occur
in cryogenic environments, called thermo-acoustic oscillations
(TAO). For SPIRAL2, TAO can be troublesome for several reasons and can
have consequences on our ability to reliably operate the accelerator
at the required energies. These oscillations are not new for a
superconducting LINAC. In fact, several studies have been reported on
oscillations occurring in individual accelerator components (cryostats and
cryogenic vessels) or in a whole LINAC \cite{fuerst1990,
  castellano1996, campisi20054, lobanov2017}. However, this phenomenon remains difficult to measure, study, understand and suppress in complex environments, where multiple thermoacoustic resonators may be excited at the same time.
This paper reports systematic investigation of thermo-acoustic oscillations distributed over the SPIRAL2 LINAC. A first part describes the system under consideration and its critical components, as well as the detailed experimental setup and the different solutions that have been considered to solve the problem. A second part shows measurements of the studied phenomena over several years and under different configurations. It investigates localised and non localised resonance phenomena in the LINAC. It also explores different damping solutions and their effect on the cryogenic operation. Finally, a conclusion and future prospects close this study.


\section{\label{sec:2}System description}

\subsection{\label{sec:2.1}System and devices under consideration}
From the cryogenic perspective, the heart of the SPIRAL2 LINAC is made of 19 cryostats (called cryomodules or CM). These cryomodules are spread along the beam line and comprise the accelerating structures : superconducting radio-frequency (RF) quarter wave resonating (QWR) cavities. They are connected to a valves box. The latter ensures the connection to a common cryodistribution line and feeding the cavities with near atmospheric pressure liquid helium at $\sim 4~K$ and the thermal screens with 14 bar 60 K helium gas. The SPIRAL2 cryoplant, centred around an Air Liquide Helial LF cold box provides the necessary cooling power. It can supply 120 g/s (1100 W at 4.3 K) \cite{ghribi2017, ghribi2017_2}.  It is worth mentioning here that the cryomodules are of two kinds. 12 of them enclose a single cavity and are called type A. The other 7 enclose two cavities each and are called type B. The valves boxes that manage the fluids of these cryomodules also have some geometrical differences. One of the main roles of the cryogenic system is to maintain stable conditions during operations such as to keep all the cavities at a stable and uniform temperature (plunged in liquid helium) and with pressure variation within the requirements\footnote{5 to 7 mbars depending on the cryomodules}. If the liquid helium level drops in a cryomodule’s phase separator, there is a risk that the corresponding cavity quenches, ie. loses its superconducting state. If the pressure in the phase separators varies too much and too quickly, the efforts applied on the cavities surface result in an elastic deformation of its shape. That changes its impedance in a way that the cavity is not matched anymore to its frequency of operation. There are of course a number of corrections applied to the RF impedance or phase changes. For instance, the Low Level Radio Frequency (LLRF) system feeding the RF power to the cavity can manage a certain bandwidth correction at the cost of some RF power. This correction is fast (high frequency) and limited to small variations. The frequency tuning system can manage slower corrections (more than one second), adapted to large variations occurring within its range of operation. The third and final way to limit the impedance fluctuations of the cavities is to control the pressure in the phase separators. This stringent requirement for a bi-phasic cryogenic operation has led to several model-based developments of the manner input and output valves can be controlled \cite{Vassal_2019, Bonne_2020}. However, using a model-based control relies on the knowledge of the internal state of system. In our case, the appearance of thermo-acoustic oscillations brought in some additional dynamics that resulted in errors in the prediction of the behaviour of our system. Among other effects, we observed additional heat load, cavities de-tuning and, overall, unstable thermodynamic behaviour.

\begin{figure}[hbt]
  \includegraphics[width=.5\textwidth, right,trim={7.5cm 1.5cm 6cm 2cm},clip]{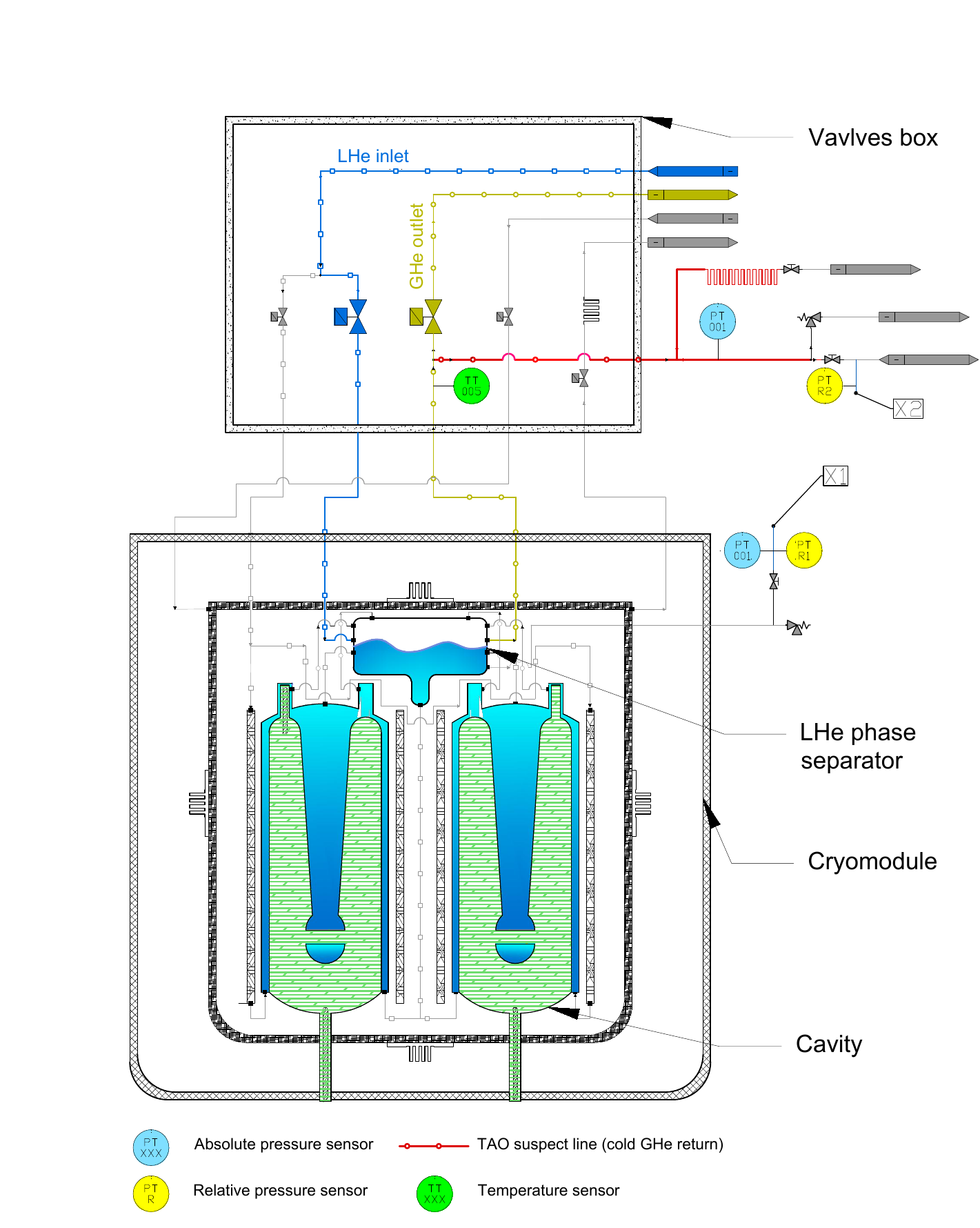}
  \caption{P\&ID of a type B cryomodule and its valves box. The red line is the low pressure return line from the LHe bath to the room temperature passive heater. Yellow circles are the rapid piezoelectric acoustic pressure sensors. Blue circles are the absolute pressure process sensors. The green circle is the position of the temperature sensor on the return line.}\label{setup_tao}
\end{figure}

\subsection{\label{sec:2.2}Root cause investigation}

\begin{figure}[hbt]
  \includegraphics[width=.5\textwidth, right]{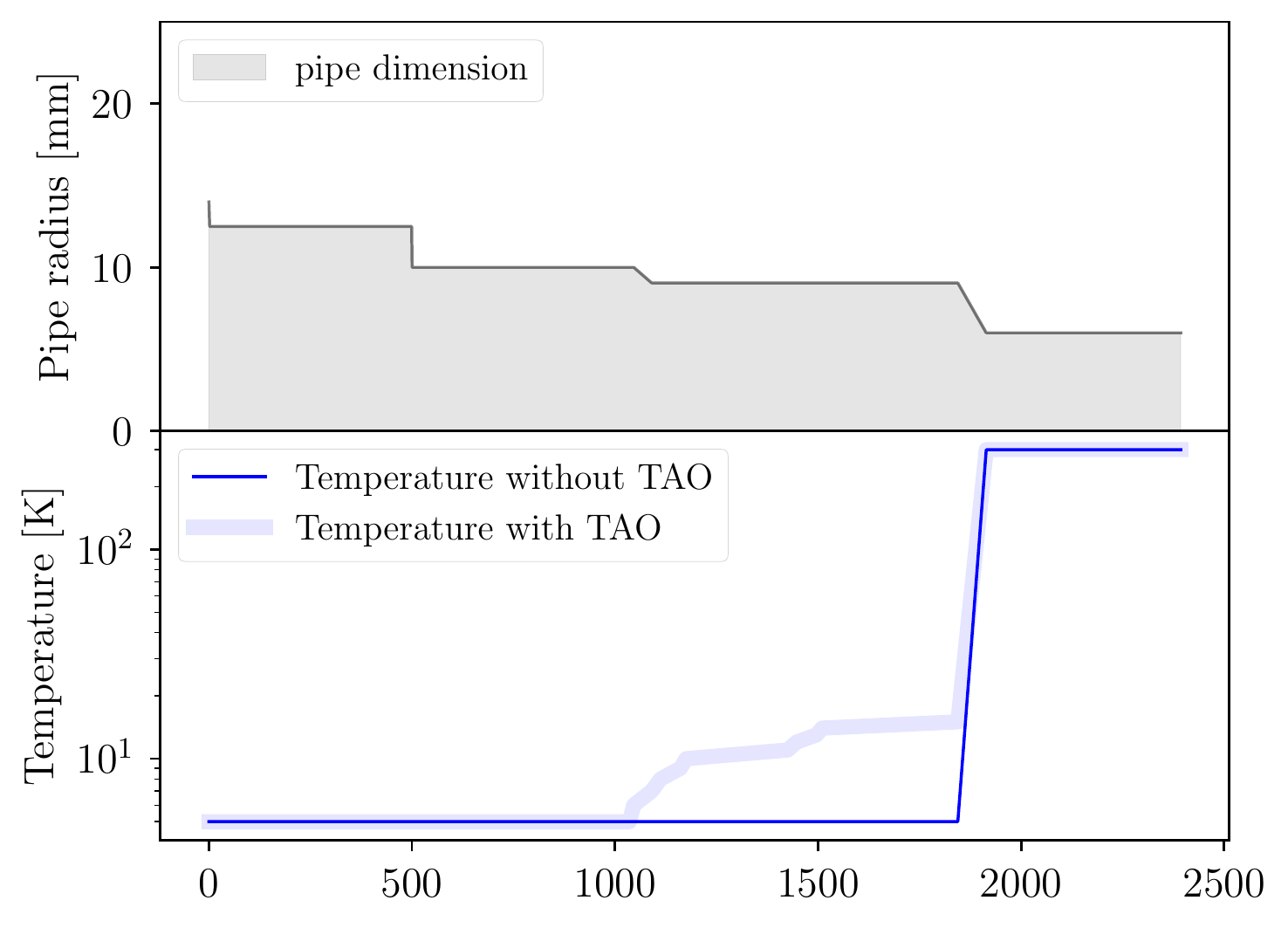}
  \caption{Main process cold return line to output purge line (red line in figure \ref{setup_tao}): pipe geometric parameters and temperature along its length with and without thermo-acoustic oscillations.}\label{fig:pipe}
\end{figure}

The first detection of acoustic oscillations in the SPIRAL2 LINAC were done thanks to RF measurements\cite{GHRIBI2020103126}. These measurements showed amplitude modulations of the transmitted and reflected RF signal. Modulation frequencies were stable but ranged from 4 to 6 Hz depending on the cavities positions. Joint piezoelectric pressure measurements of the cavities liquid helium phase separators showed direct correlations (see \ref{sec:2.3} for the experimental setup). The first root cause investigations were done with tri-axial accelerometers. Thanks to them, all external mechanical vibrations were ruled out. Accelerometers showed vibrations along a single axis corresponding to the direction of the cryo-fluids distribution. Vibration amplitudes increased at one of the valves-boxes room temperature ends. Although not frozen, the identified room temperature port was slightly colder than other room temperature ends. This identified port was used for the purge and pressure measurements of the main process return line (saturated helium). Incidentally, it was noticed an abnormally high temperature of the cryomodules helium return line. At the same time, measurements of static heat loads of the cryomodules showed values remaining in good agreement with the specifications. All these inputs led us to identify the line where thermo-acoustics were likely to appear. Figure \ref{setup_tao} shows
a schematic view of a cryomodule and its valves-box with the identified line where TAO occur colored in red. Figure \ref{fig:pipe} shows the changes of the line geometry and its temperature along its length. These data are used as inputs for the Rott’s study of the likelihood of appearance of TAO given these geometries and temperatures (see \cite{ghribi2021cryogenic}).

\subsection{\label{sec:2.3}Experimental setup}

\begin{figure}[h]
  \centering\includegraphics[width=.4\textwidth, right,trim={3cm 3cm 4cm 4cm},clip]{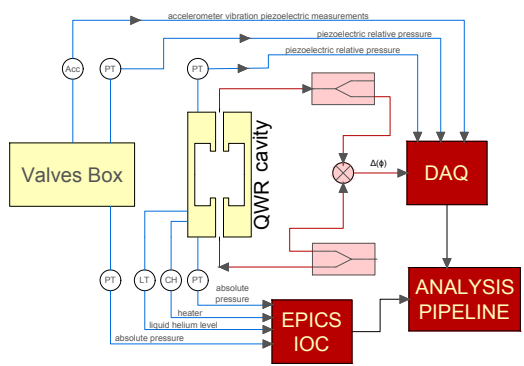}
  \caption{Block diagram of the 2018 measurement setup.}\label{schema_bloc2}
\end{figure}

Two kinds of measurements were performed in order to better understand and suppress these thermo-acoustic oscillations. The first one was a set of measurements targeting the characterization of amplitudes and frequencies of the oscillations without any modification of the system and under nominal operating conditions (pressure, heat load, liquid helium level). It included simultaneous measurements of absolute pressures, acoustic pressures, RF phase shifts, liquid helium level, and heaters powers at different locations of a target cryomodule. A block diagram of such a setup is shown in Figure \ref{schema_bloc2}. We measured both the absolute and the dynamic acoustic pressure directly in the cryomodule phase separator and in the matching valves box return line. We used piezoelectric sensors (PCB 113B28) for acoustic measurements and metallic process isolating diaphragm sensors (Cerabar PMP71) for absolute pressure measurements. The same setup (Figure \ref{schema_bloc2}) was used to extract the phase shift between the RF input and output signals. A National Instrument Compact DAQ centralized the fast acquisition with 3 channels 2.8 kS/s/ch 24-Bit analog modules for the IEPE signals (Integrated Electronics Piezo-Electric) and a universal analog module for the RF signal. The NI DAQ was driven by an external laptop running Labview. Other data such as absolute pressures, heater power, liquid helium level and temperatures were measured through our regular Programmable Logic Controllers (PLC) and archived with an Experimental Physics and Industrial Control System (EPICS) Input/Output Controller (IOC). A Python analysis pipeline assembled fast and slow acquisitions together with other correlation factors and a common clock.

\begin{figure}[h]
  \centering\includegraphics[width=0.45\textwidth]{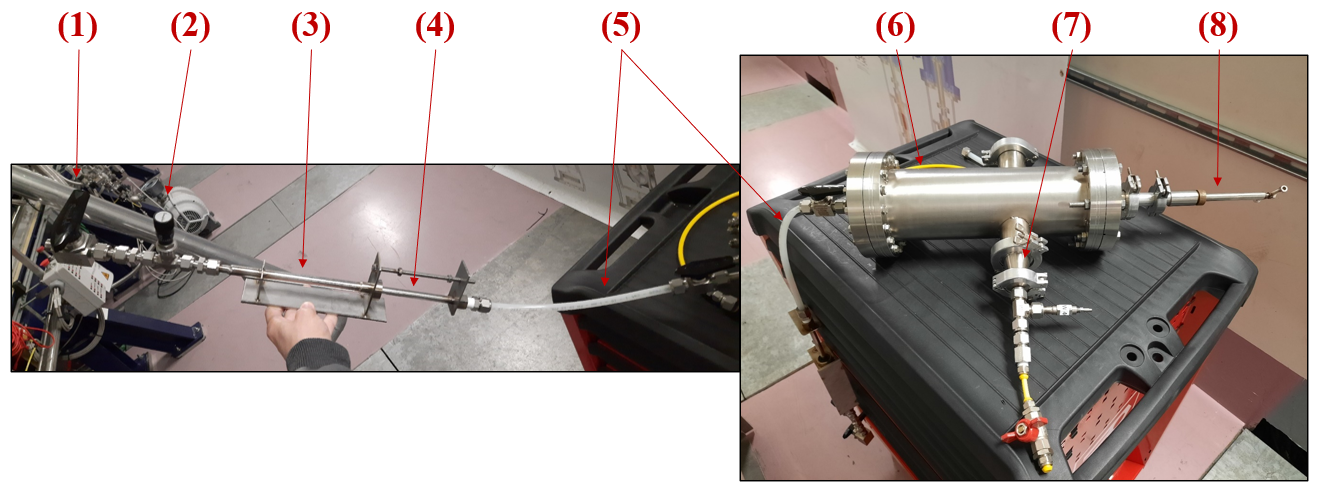}
  \caption{The RLC resonator: (1) High pressure ball valve, (2) Micrometer needle valve, (3) Fix inductance, (4) Variable inductance, (5) Connection tube, (6) Capacitance volume, (7) Purge access, (8) Capacitance piston.}
  \label{RLC bench}
\end{figure}

The second set of measurements used an acoustic resonator, connected to the warm side of the LINAC. This resonator was made up of three main adjustable acoustic elements: a resistance, an inductance, and a capacitance, therefore called “RLC resonator” using an electro-acoustic analogy. The first purpose of the bench was to identify the best suitable configurations to efficiently damp the TAO in the LINAC. The second purpose was to estimate the acoustic impedance of the system under investigation for further studies and developments. The same acquisition system, although limited to piezoelectric pressure measurements was used.
In the resonator, the resistive element is a micro-metric needle valve, the inductive element a small diameter tube (10mm diameter) of variable length, and the capacitive element (compliance) a large diameter pipe (100mm diameter) of variable length. The tuneability of the compliance was achieved by use of a gas-tight piston (see Figure \ref{RLC bench}).

According to the electro-acoustic analogy \cite{Swift}, the inductance and the compliance of a channel can be calculated using the following two equations respectively:
\begin{equation}
 L = \frac{\rho_m \Delta x}{A}
 \label{eq1}
\end{equation}
\begin{equation}
 C = \frac{V}{\gamma p_m}
 \label{eq2}
\end{equation}
where $\rho_m$ is the mean gas density, $\Delta x$ the channel length, $A$ the channel cross-sectional area, $V$ the channel volume, $\gamma$ the gas heat capacity ratio, and $p_m$ is the gas mean pressure.



\section{\label{sec:4}Observations and measurements}

\subsection{\label{sec:4.1}Full kernel distributions of cross-coupled and decoupled behaviours}

While it might seem obvious, for an isolated system, that thermoacoustic oscillations occur because of local conditions, the answer is not clear for coupled or connected systems. For the LINAC, we are in the complex case of interconnected cryogenic clients with several room temperature ports with cold ends. It is therefore unclear wether the amplitude and frequency dependance of the oscillations is dominated by local effects or global effects. Observed transient pressure fluctuations with sudden changes in the frequency of the TAO could be caused by such interconnections. 

\begin{figure}[hbt]
  \centering\includegraphics[width=.5\textwidth, right,trim={0cm 0cm 0cm 0cm},clip]{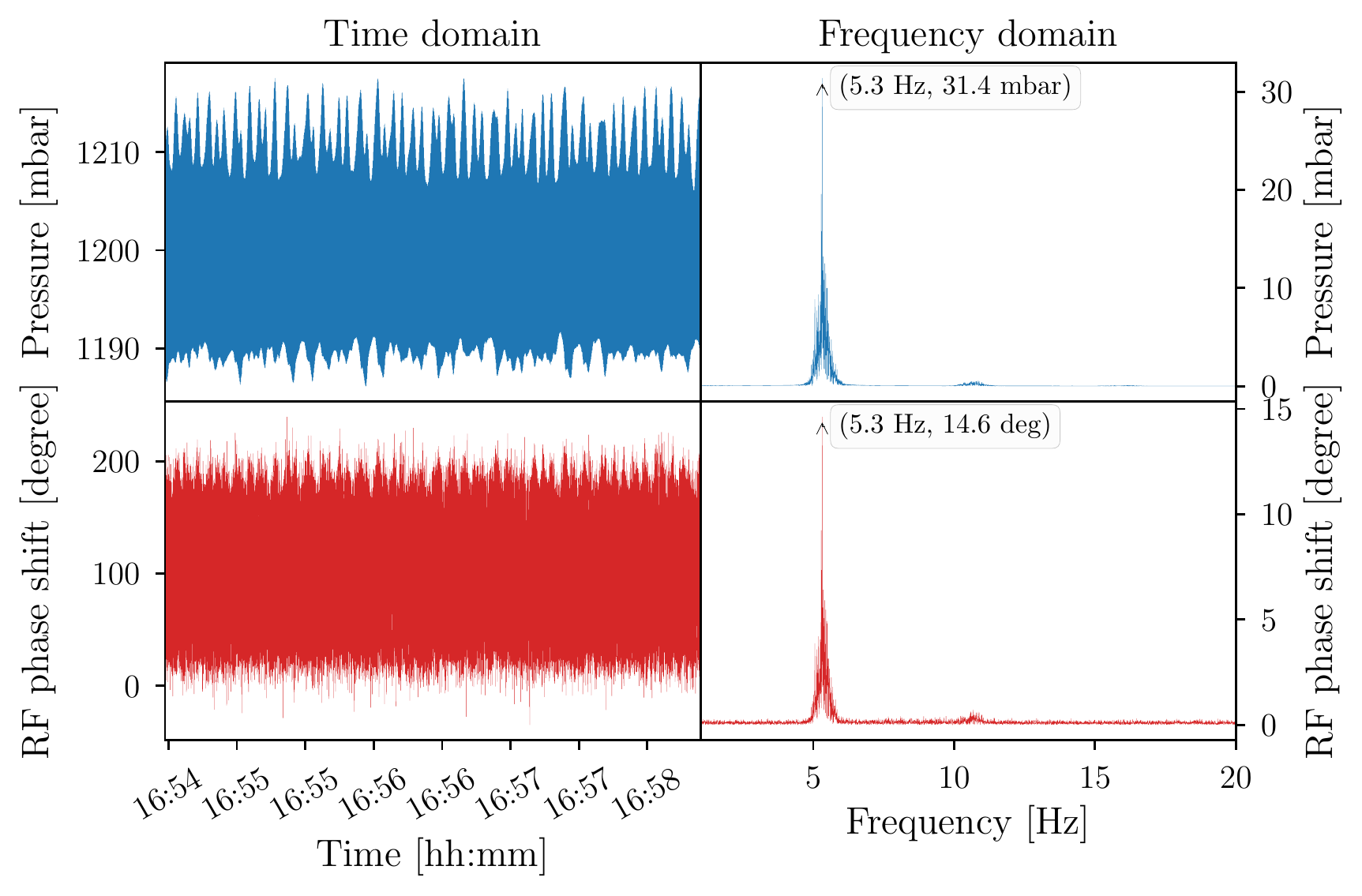}
  \caption{Example of time and frequency domain data extracted and calculated with the pipeline analysis for CMA11 [2018-11-11]. Frequency domain amplitudes shown here are peak-peak amplitudes.}\label{timef}
\end{figure}

\begin{figure}[hbt]
  \centering\includegraphics[width=.5\textwidth, right,trim={0cm 0cm 0cm 0cm},clip]{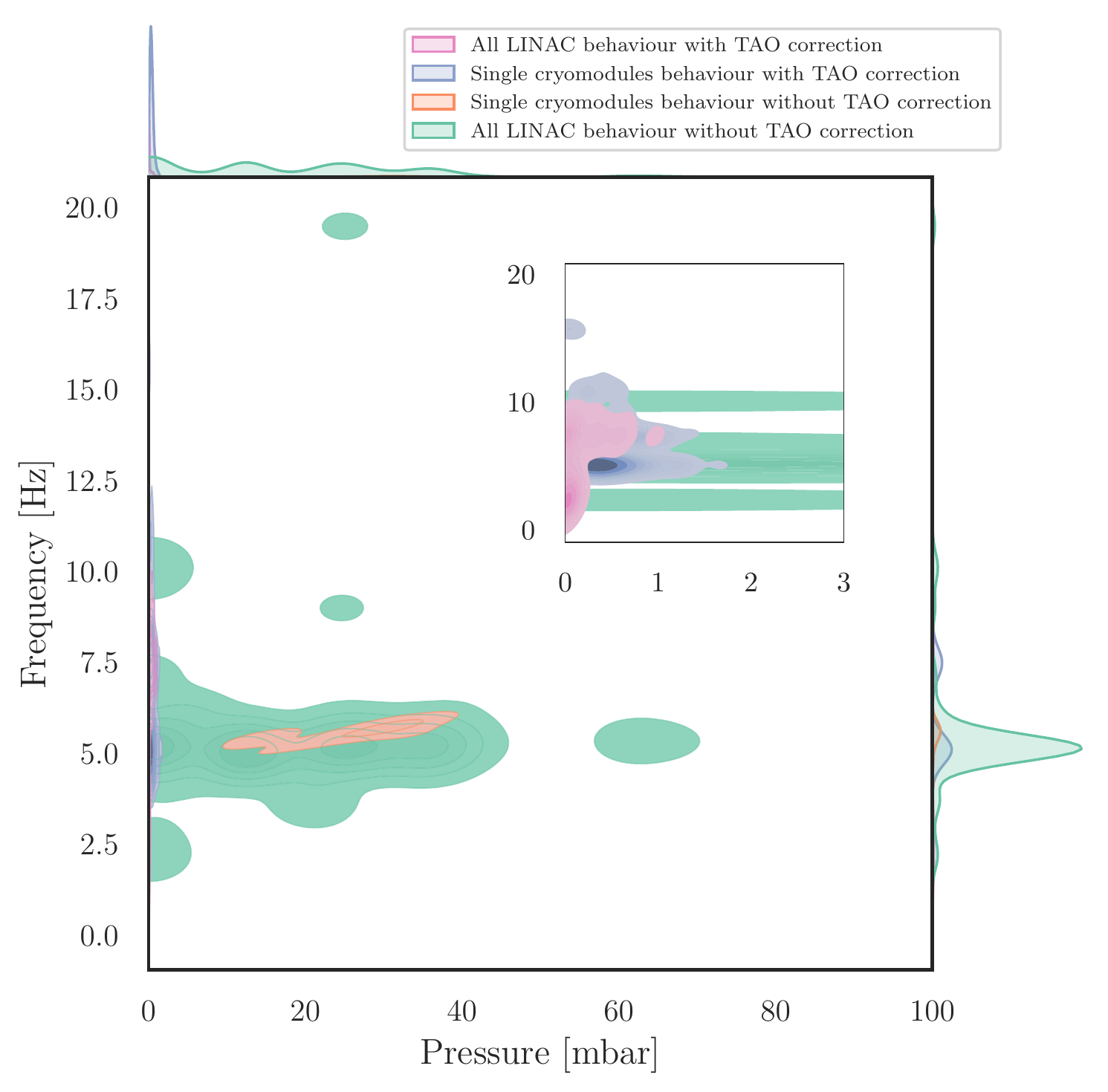}
  \caption{Amplitudes and frequencies kernel distributions of detected TAO oscillations for different configurations.}\label{kernel}
\end{figure}

To have a better picture, we deployed the setup described in Figure \ref{schema_bloc2} all along the LINAC. We therefore deployed 19 piezoelectric acoustic pressure sensors (one for each cryomodule) in the LINAC and one on the main return line. Acquisition was made by the same DAQ  previously described with seven 24-bit analog modules controlled with a Labview program. As previously, all fast acquisition data were treated with a python pipeline analysis that combined PLC slow sensors and NI DAQ fast sensors. This time, the pipeline allowed automatic peak extraction and TAO detection. This broaded our view of the phenomena occurring in the LINAC.  All data were gathered in time domain. The reference of a piezo-electric pressure sensor is always zero which means that the correct amplitude is found by offsetting the relative data (voltage output of the piezo-electric sensors) by the measured absolute pressures (output of the process pressure sensors). The amplitude of the oscillations was found by enveloppe calculations within the considered time window using its Hilbert transform. For the frequency peak detection, we first applied a Fourier transform on the relative data within a time window of 4 minutes in order to have a high resolution. We then applied a high pass filter to avoid the 1/f noise below 1 Hz and a low pass filter to avoid high order harmonics. We finally computed the centroid of the resulting spectrogram to extract the frequency peak. An example of a time and frequency domain extracted data is shown in Figure \ref{timef}. 
Thanks to the extracted data, Parzen-Rosenblatt kernel density estimations were computed for both the frequencies and the amplitudes of TAO for every considered configuration. Figure \ref{kernel} depicts the analysis of 10,787 datasets. The TAO correction used here is the short circuit line correction described in \ref{sec:sbp}. The label \emph{All LINAC behaviour without TAO correction} refers to simultaneous TAO measurements in all the LINAC without any TAO correction. The label \emph{Single cryomodules behaviour without TAO correction} refers to simultaneous TAO measurements with TAO correction applied to all LINAC except one cryomodule. The position of the cryomodule which has no active TAO correction is spanned all over the LINAC. The resulting data therefore represents the oscillations of all cryomodules in a configuration where their single behaviours dominate. We can easily see in Figure \ref{kernel} that both single cryomodules and the overall LINAC resonate around frequencies comprised between 2.5 and 10 Hz. When only one cryomodule resonates (label ``Single cryomodules behaviours without TAO correction'' of Figure \ref{kernel}), the frequencies of oscillations stabilise with a narrow bandwidth. This seems to point that the resonance frequencies are dominated by local effects and therefore that these phenomena occur in the same physical region of the measured cryomodule-valves box pair. When all cryomodules resonate at the same time (label ``All LINAC behaviour without TAO correction'' of Figure \ref{kernel}), the frequencies and amplitudes of the oscillations are distributed along a wider range of values. This indicates that the overall spread of both frequencies and amplitudes are dominated by cross-couplings. This cross-coupling appears critical only when multiple TAO are active at the same time.
When no TAO is active (zoom window in Figure \ref{kernel}), resonance amplitudes stay below 2 mbars. In this latter configuration, single cryomodules resonances seems to emphasise to frequencies (4 and 6 Hz). Cross-couplings in this same configuration flattens this frequency behaviour towards lower frequencies.

\subsection{\label{sec:2.4}Applied damping solutions}

\subsubsection{\label{sec:sbp}Short, buffer and piston}
Prior to the detailed system study, fast, efficient and non-intrusive solutions had to be found quickly to damp the oscillations and allow for the commissioning of the SPIRAL2 LINAC. Several solutions documented in \cite{GU1992194, ditmars1965, CHEN1999843, LUCK1992703} were investigated :

\begin{itemize}
\item Short circuit between the phase separator vapour sky and the return line : Here we linked the two ports X1 and X2 (see Figure \ref{setup_tao}) with several pipes of several length and cross-sections.
\item Buffer : here we connected several buffers of different volumes to the port X2.
\item Piston  : here we inserted a piston in the port X2 and we monitored the behaviour of the system for several insertion depths.
\end{itemize}

For every tested solution, we spanned all operating conditions to determine the most suitable solution for our case. In order to compare the results, we used a damping efficiency criterion defined as :
\begin{equation}
	\zeta = \frac{P_{bath}^{off} - P_{bath}^{on}}{P_{bath}^{off}} \label{eq:zeta}
\end{equation}
where $P_{bath}^{off}$ is the amplitude of the pressure oscillations in the liquid helium phase separator with no TAO correction and $P_{bath}^{on}$ is the amplitude of the pressure oscillations with the considered TAO correction. $\zeta\rightarrow 1$ would give us a good damping while $\zeta\rightarrow 0$ will indicate almost no damping effect.

\begin{figure}[h]
  \centering\includegraphics[width=.5\textwidth, right,trim={0cm 0cm 0cm 0cm},clip]{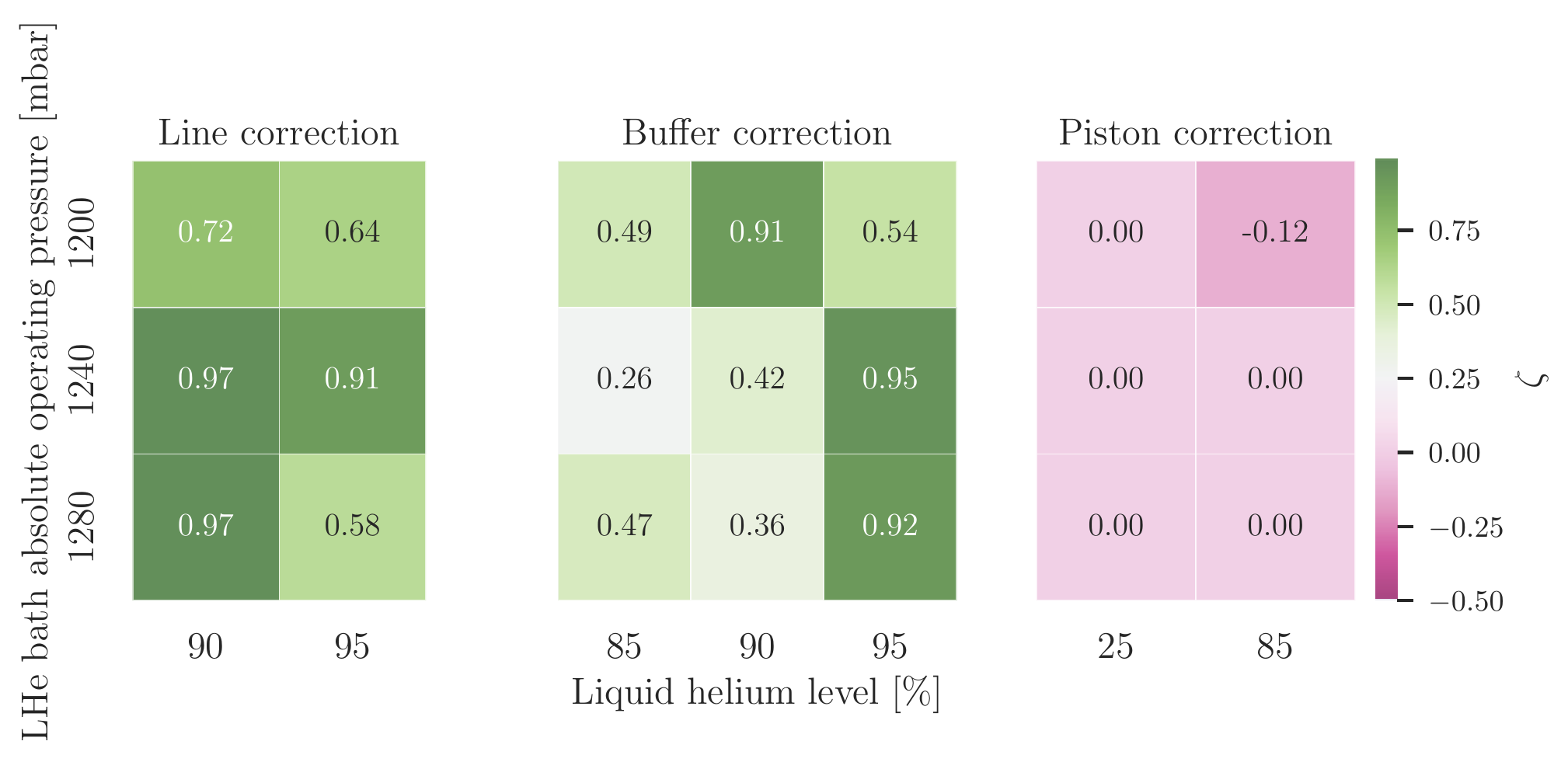}
  \caption{TAO correction efficiency $\zeta$ as a function of the pressure (PT001) and the liquid helium level (LT200) for the three considered experimental corrections : line (vapor sky short circuit), buffer and piston for cryomodules CMA05 and CMB03.}\label{eff_taoc}
\end{figure}

For every applied correction, we did see an effect on TAO damping but no total suppression was achieved. An example of the efficiency reached with every correction is shown in Figure \ref{eff_taoc} for different pressures and liquid helium operating conditions. It appeared that the most efficient solution for every case is the line short circuit correction. This solution was efficient enough to be deployed in all LINAC. The TAO correction efficiency reached 0.97 for some cases, damping the oscillations amplitudes to acceptable values. Experiments showed that high amplitude thermo-acoustics appeared when pressure difference between port X1 and X2 reached 0.1 mbar. the The short circuit line was sufficient to recover this pressure unbalance. However, the flow rate was so important for some cryomodules locations that it froze part of the line and the upper neck of the cryomodule or resulted in some condensation at the same locations. We then deployed the short circuit line solution to all cryomodules. The lines were terminated by a on/off hand valve at one end, a micro-metric hand control valve at the other hand and pressure security valve in between. The micro-metric valve limited the flow in the correction line in order to avoid water condensation or ice. The on/off valve was meant to suppress any flow through the correction line. This was useful especially when cooling down the LINAC. The same valve was also used to de-activate or activate the TAO correction at will for a more thorough investigation of TAO amplitudes and frequencies cross-couplings (see subsection \ref{sec:4.1}).

\subsubsection{The special case of the RLC resonator}

\begin{figure}[h]
  \includegraphics[width=0.46\textwidth]{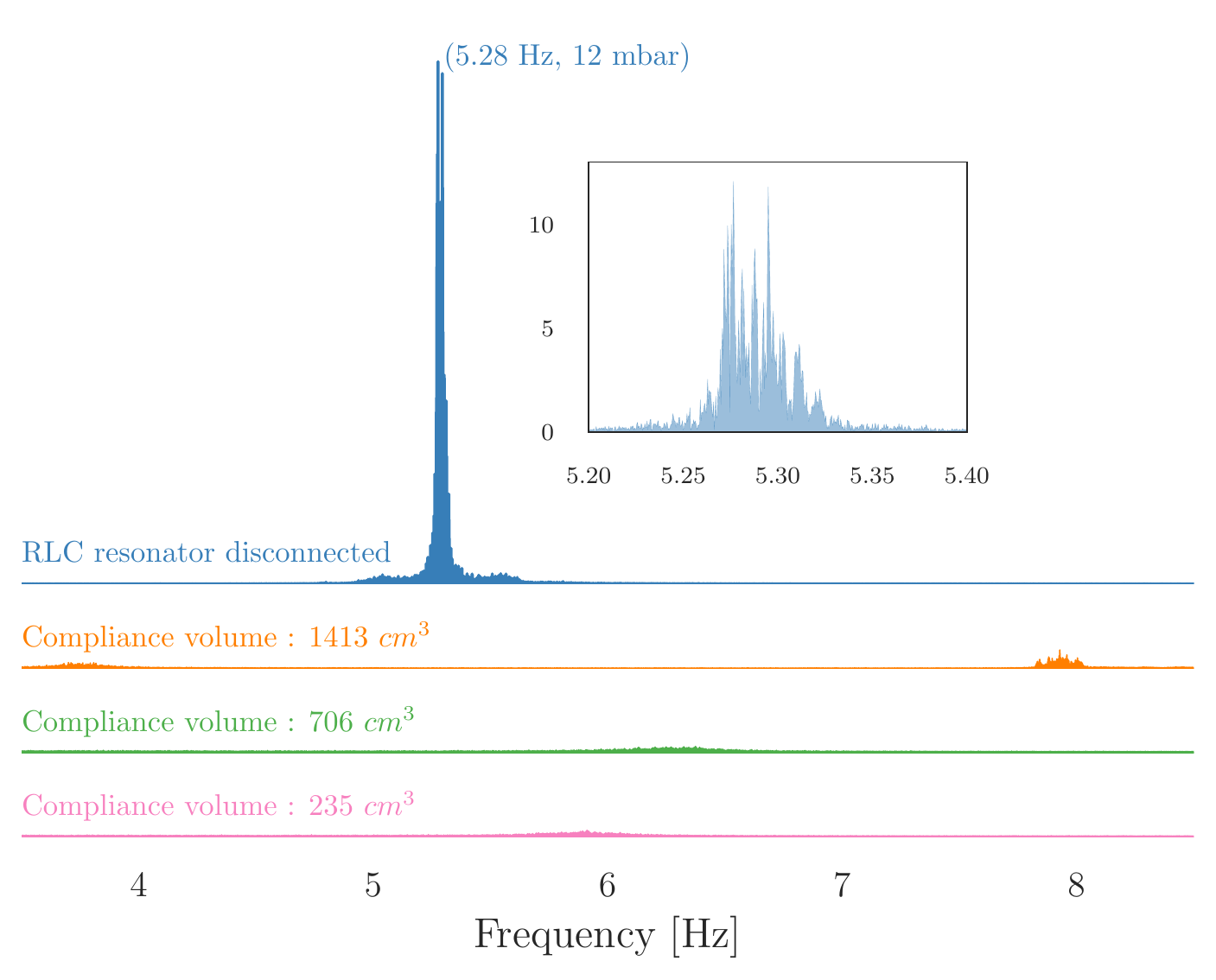}
  \caption{Pressure oscillations measured in the liquid helium phase separator of cryomodule CMA12 for four different configurations : 1. No RLC resonator connected (blue) | 2. RLC resonator with a $1413~cm^{3}$ compliance volume | 3. RLC resonator with a $706~cm^{3}$ compliance volume | 4. RLC resonator with a $235~cm^{3}$ compliance volume. The plots use the same xy axis but with a shifted y axis for increased visibility.}
  \label{CMA12}
\end{figure}

The RLC resonator was tested on two type A cryomodules, which had more instabilities in comparison with type B ones. The resistance (the micrometric valve) demonstrated to be the dominant effect on the damping of the pressure perturbation, meaning that we were able to effectively damp the oscillations whatever the capacitive and inductive volumes were. The correction efficiencies were comparable to those found with the short circuit line (see \ref{sec:2.4}). 
For example, in figure \ref{CMA12}, we see a comparison between the level of oscillations observed on CMA12 without the RLC bench and for three different volumes for its compliance (the inductance was maintained at a constant position). As we can see, the amplitude of oscillations was effectively decreased (from 12 mbar to about 0.5 mbar) without tuning the inductance.

In order to be able to investigate the effect of a tuneable compliance and inductance on TAO damping, the resistance (i.e. micro-metric valve) was suppressed. To study the efficiency of the LC bench, two parameters have been considered. The first one is the damping efficiency criterion $\zeta$ defined in eq. \ref{eq:zeta}. The second is the acoustic impedance, respectively expressed for the tuneable inductance and compliance by : 
\begin{equation}
 Z_L = i\omega L
 \label{eq3}
\end{equation}
\begin{equation}
 Z_C = 1/{i\omega C}
 \label{eq4}
\end{equation}
where, $i$ is the imaginary symbol and $\omega$ is the angular frequency of resonance. The unit of the specific acoustic impedance is the $Pa.s/m^3$ or the acoustic Ohm $\Omega_a$.

\begin{figure}[h]
  \includegraphics[width=0.46\textwidth]{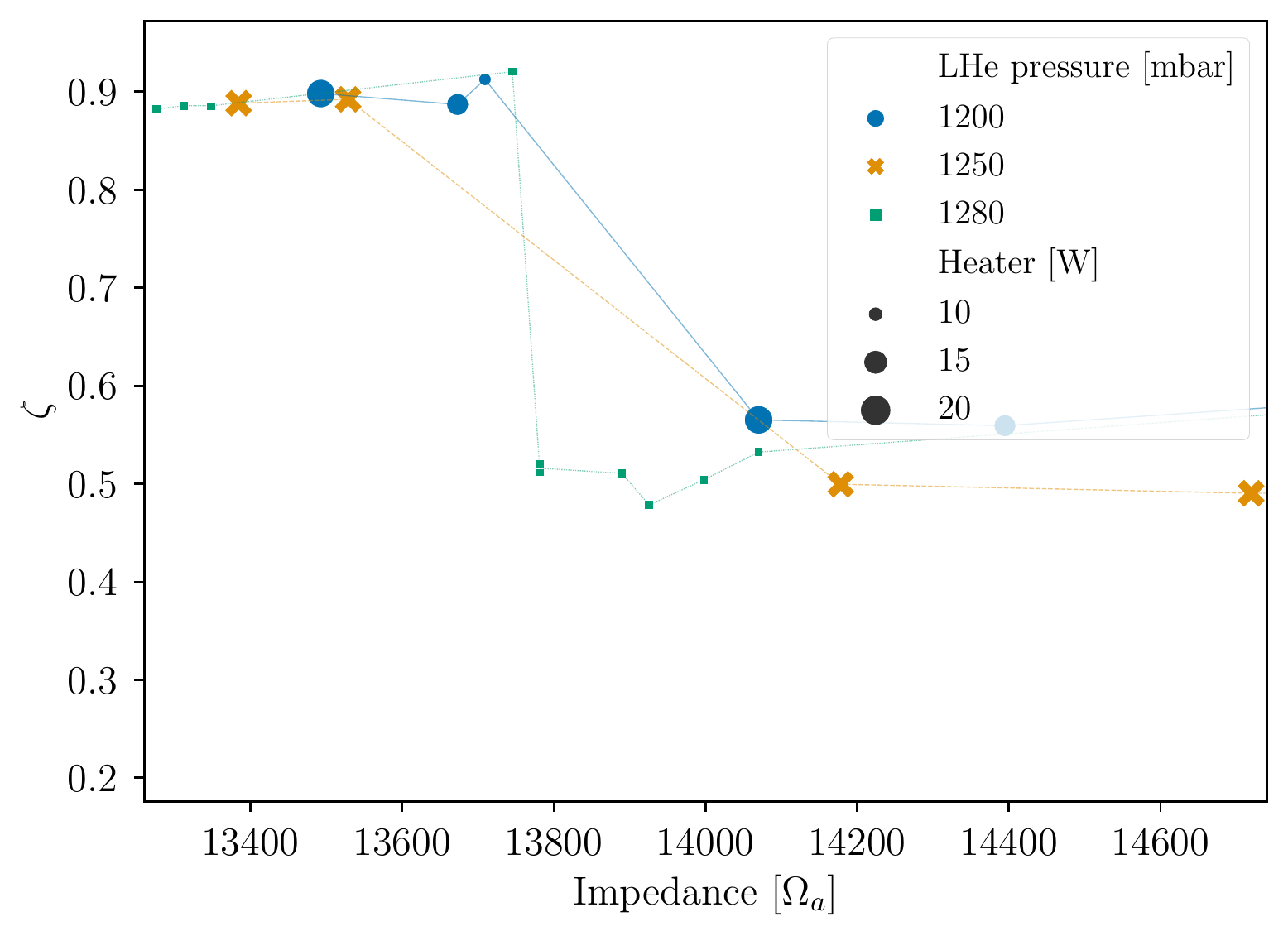}
  \caption{Damping efficiency criterion $\zeta$ at CMA04 position as a function of the corresponding LC resonator impedance for different LHe bath pressures and heater powers.}
  \label{DR vs imp}
\end{figure}

Several inductance lengths and compliance volumes have been investigated at different LHe bath pressures and heater powers.
The resulting damping efficiency criteria at CMA04 position are plotted as a function of the impedance of the LC resonator in figure \ref{DR vs imp}. The latter shows that $\zeta$ suddenly drops from $\sim 0.9$ to $\sim 0.6$ between $13.8~k\Omega_{a}$ and $14~k\Omega_{a}$. Although limited to one cryomodule, these data constrain the region of investigation for an efficient damping under different operating conditions. Consequently, a chosen configuration of the resonator within the identified impedance region was deployed separately on a sample of different cryomodules across the LINAC. The results showed resonances with amplitudes below 0.5 mbar but with a different frequency behaviours for types A and B CM. Types A showed sharp low frequency peaks between 4 Hz and 7 Hz while types B experienced more widely spread frequencies from 4 Hz to 100 Hz. The simultaneous deployment of the resonator on all cryomodules is still to be made in order to investigate possible cross-couplings and how critical they could be on operation.


\subsubsection{\label{sec:sideeffects}TAO correction side effects on operation and control}
\begin{figure}[hbt]
  \includegraphics[width=.5\textwidth, right,trim={0cm 0cm 0cm 0cm},clip]{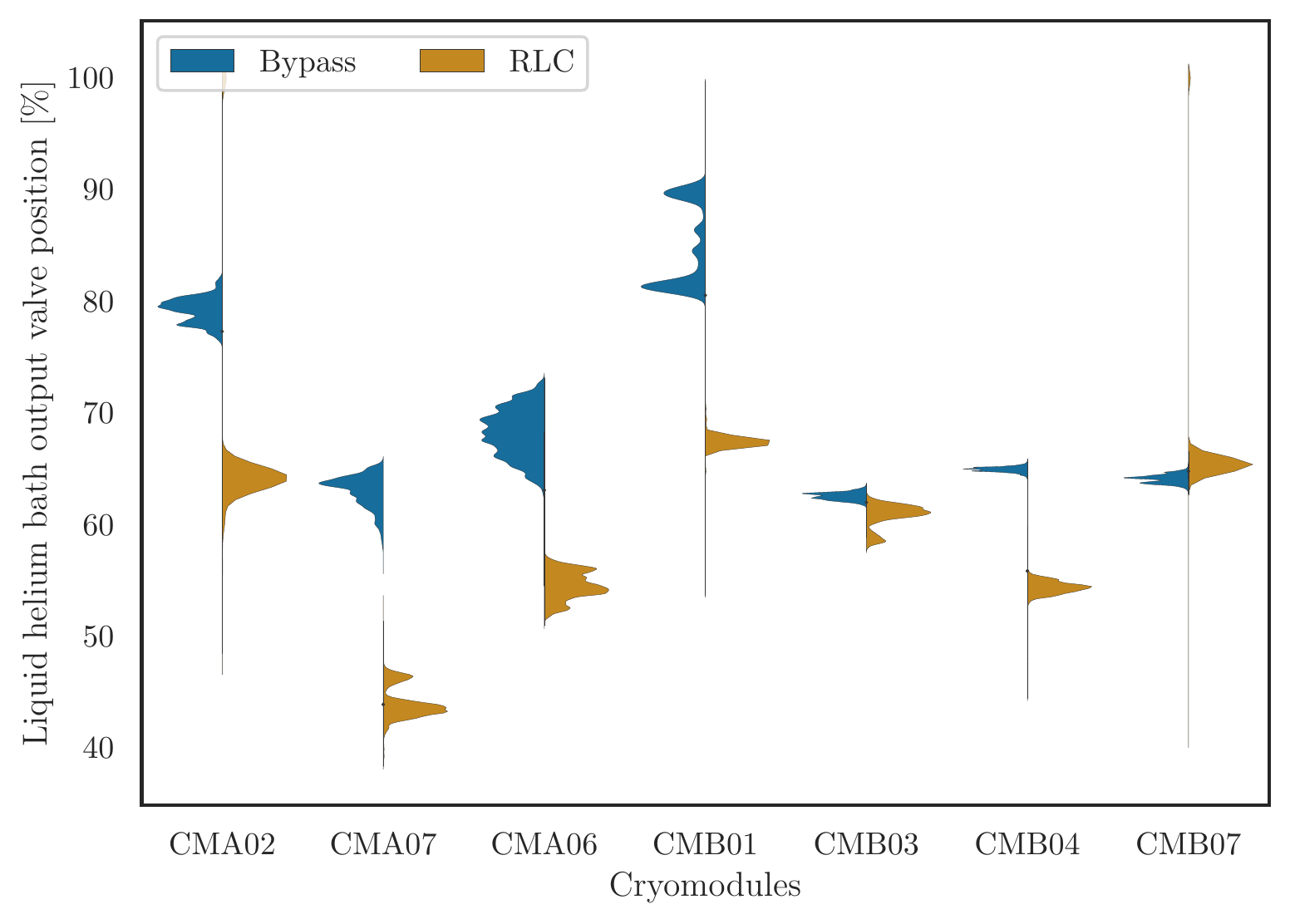}
  \caption{Split-violin plot distributions of the measured positions of the main process return valves for 7 cryomodules [measurements of 2021-10-11].}\label{fcv005bp}
\end{figure}

Cryogenic thermo-acoustics are known to be the source of multiple problems in a superconducting LINAC. These can for example cause room temperature ports freezing, unstable liquid helium level readout and unstable liquid helium bath pressure. However, getting rid of these oscillations can, by itself, be the source of other \emph{side effects} on the cryogenic operation of the LINAC. In the case of SPIRAL2, the first emergency solution that has been applied, called \emph{bypass line} or \emph{short-circuit line}, helped balance the pressure difference between two critical points, removing the condition of appearance of thermo-acoustics. This implied a variable helium flow rate that bypassed the main isolated process return line linking the cryomodule to the valves box. This behaviour can be seen as a variable impedance of the main return line and an added low density warm helium return to the main saturated helium return. As a consequence, the outgassing valves saw an effective flow reduction equivalent to an effective virtual reduction of the pipe diameter for the same operating condition. These phenomena appeared thanks to virtual flow observers that showed an unbalance between the input helium mass flow rate and the output helium mass flow rate. Replacing the \emph{bypass line} with the RLC resonator proved efficient in avoiding such unbalance. This can be seen in Figure \ref{fcv005bp} where the shape distributions of all output valves positions change with a noticeable drop of the mean values. This behaviour is however heavily dependent on the position of the cryomodules and their geometries. Most type B (doubles cavities) cryomodules are for instance less sensitive to thermo-acoustics and therefore show less difference between the two damping solutions considered here. Type A (single cavities) cryomodules on the other hand are very sensitive to thermo-acoustics, which makes them good indicators to the most suitable damping solution for the cryogenic operation.

\section{Conclusions}
The SPIRAL2 superconducting LINAC is a school case of cryogenic thermo-acoustics. However, with its distributed cryogenic clients and complicated control system, dealing with these oscillations in every day life operation can be more complex that it might be for a single cryostat in a laboratory test bench. To overcome these difficulties, the first step has been to provide the accelerator with simultaneous eyes to detect and monitor these phenomena. The full integration of the fast acquisition system, processing and oscillations detection within the accelerator control system is a step further that is planned in the near future. Monitoring these oscillations allowed to quickly find a fast and efficient solution to continue the commissioning and operation of the LINAC. Meanwhile, modelling the thermodynamic behaviour allowed to highlight damping side effects on the cryogenic operation. A variable impedance resonator has been designed and put in place based on previous developments at partner laboratories (IJCLab and CES). The resonator was meant to study the acoustic impedance of the system and derive more efficient damping solutions without the previously noticed side effects. This role has been fully achieved and the resulting new damping solution based on the now known acoustic impedance is being designed and fabricated for a setup in the accelerator during spring 2022.

Being able to switch at will thermo-acoustic oscillations and study cross-couplings between cryogenic clients makes SPIRAL2 a perfect laboratory to study and harness these phenomena for other applications such as traveling wave thermo-acoustic power generation.

\subsection*{Acknowledgements}
This work has been funded by "Region Normandie" as well as the city of Caen, CNRS and CEA. We would like to thank all contributors from CEA-IRFU, CNRS-IJC Lab and GANIL without whom this paper would not have been possible. We also thank F. Bonne and P. Bonnay (DSBT/CEA) for the Simcryogenics library that is being used for the model base control of the cryogenics. We thank F. Millet (DSBT/CEA) for useful discussions on liquid helium level sensors shielding. We finally thank D. Longuevergne and M. Pierens from IJC Lab for kindly borrowing us the first fast sensors and acquisition system for vibrations investigations in 2017 and useful discussions on setting up the first experiment.

\subsection*{Data availability}
Raw data were generated at the SPIRAL2 facility. They are not publicly available at the moment due to CNRS/CEA policy restrictions. The data that support the findings can be available from the corresponding author upon reasonable request and with permission.

\bibliography{ref}

\end{document}